\documentstyle[11pt,newpasp,twoside,epsf]{article}
\def\edcomment#1{\iffalse\marginpar{\raggedright\sl#1\/}\else\relax\fi}
\reversemarginpar

\begin{document}
\title{Be stars in X-ray binary systems}
 \author{M.J.Coe}
\affil{Department of Physics and Astronomy, The University,}

\begin{abstract}

This paper will review the status of our observations and
understanding of Be stars in X-ray binary systems. In virtually all
cases the binary partner to the Be star is a neutron star. The
circumstellar disk provides the accretion fuel and hence stimulates the X-ray
emission, whilst the neutron star provides a valuable probe of the
environment around the Be star. The results coming from studies of
such systems are helping in our understanding of the Be phenomenon.

\end{abstract}

\section{Introduction}

The objectives of this paper are :

\begin{itemize}

\item To explain what High Mass X-ray Binaries (HMXBs) are, and, in
particular, Be/X-ray binary systems.

\item To discuss the general observational characteristics of the group.

\item To present examples of how work on such systems has furthered
our knowledge of Be stars.

\end{itemize}

\section{General properties}

The Be/X-ray systems represent the largest sub-class of massive X-ray
binaries.  A survey of the literature reveals that of the 96 proposed
massive X-ray binary pulsar systems, 67\% of the identified systems
fall within this classes of binary.  The orbit of the Be or supergiant
star and the compact object, presumably a neutron star, is generally
wide and eccentric.
X-ray outbursts are normally
associated with the passage of the neutron star through the
circumstellar disk.  The optical star exhibits H$\alpha$ line emission
and continuum free-free emission (revealed as excess flux in the IR)
from a disk of circumstellar gas.

The physics of accretion-powered pulsars has been reviewed previously
(e.g. White, Nagase \& Parmar 1995, Nagase 1989, Bildsten et al
1997).  This paper will concentrate on the optical and IR
observational properties of these systems and will address what such
observations have contributed to our understanding of Be star
behaviour.

\pagebreak 

X-ray behavioural features of these systems include:\\
-- regular periodic outbursts at periastron called Type I;\\
-- giant outbursts at any phase probably arising from a dramatic
expansion of the circumstellar disk described as Type II;\\
-- ``missed'' outbursts frequently related to low H$\alpha$
emission levels (hence a small disk), or other unknown reasons (eg
perhaps centrifugal inhibition of accretion (Stella, White \& Rosner 1986);\\
-- shifting outburst phases (see below).\\

Progress towards a better understanding of the physics of these systems
depends on a multi-wavelength programme of observations.  From
observations of the Be star in the optical and IR, the physical
conditions under which the neutron star is accreting matter can be
determined.  In combination with hard X-ray timing and flux
observations, this yields a near complete picture of the accretion
process.  It is thus vital to identify the optical counterparts to
these X-ray systems in order to further our understanding. 

Neutron stars are found in many astronomical configurations - see
Figure 1 - and Be/X-ray binaries contain the second largest group of
known neutron stars. The largest group comprises thousands of
isolated rotation-powered pulsar systems.

\begin{figure}
\begin{center}
{\epsfxsize 0.6\hsize
 \leavevmode
 \epsffile{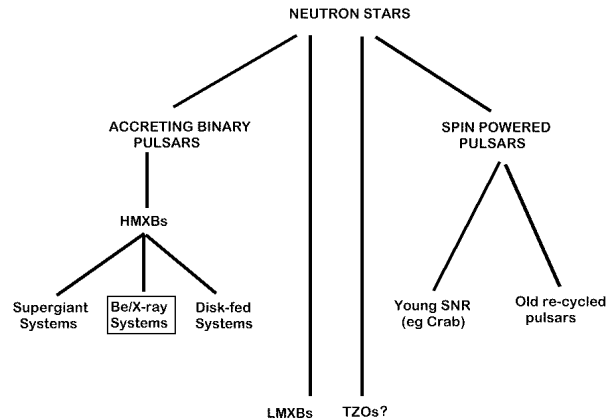}
}
\end{center}
\caption{Classification of neutron star systems}
\end{figure}

Currently (September 1999) there are about 100 known or suspected
HMXBs (see Figure 2). Surprisingly, nearly one third of all of these
sytems lie in the Magellanic Clouds. This very large fraction,
particularily noticeable in the Small Magellanic Cloud, will be
discussed below in Section 7. Currently about one third of all the
100 systems have no known optical counterpart - frequently this is
simply due to the inaccuracy of the X-ray observatory in locating the
X-ray source.

\begin{figure}
\begin{center}
{\epsfxsize 0.6\hsize
 \leavevmode
 \epsffile{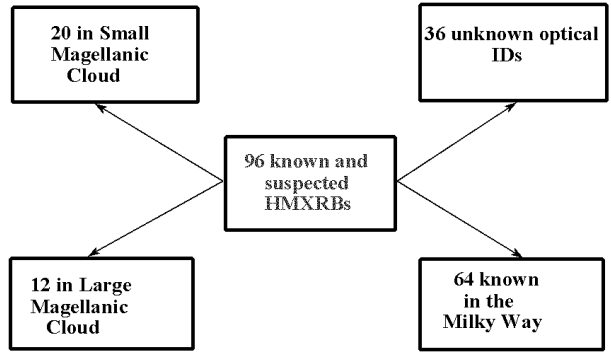}
}
\end{center}
\caption{Distribution of HMXBs}
\end{figure}

There are two main sub-groups of HMXBS - the supergiant
counterparts (normally of luminosity class I or II), and the Be/X-ray
binary systems (normally luminosity class III or V). Both systems
involve OB type stars and are commonly found in the galactic plane and
the Magellanic Clouds. They differ, however, in accretion modes with
the supergiant systems accreting from a radially outflowing stellar
wind, and the Be/X-ray binaries accreting directly from the
circumstellar disk (maybe with some limited Roche lobe overflow on
rare occasions). As a result the supergiants are persistent sources of
X-rays, whilst the Be/x-ray systems are very variable and frequently
much brighter.

Recently Reig and Roche (1999) have suggested a third sub-group of
systems; the X Persei like systems. The main characteristics of this
proposed group are: long pulse periods (typically 1000s), persistent
low X-ray luminosity ($\sim10^{34}$ erg/s) and low variability, and
rare uncorrelated weak X-ray outbursts. There are currently 5 such
objects in this new group.

A summary of the observational properties of Be/X-ray binaries is :\\
-- they were first discovered in 1974--5 (eg A1118-616 and A0535+26)\\
-- there are currently 38 optically identified systems\\
-- there are a further 23 systems proposed based upon their X-ray
characteristics\\
-- there are probably 100 -- 1000 Be/X-ray binaries in our galaxy
(Bildsten et al 1997), but maybe up to 10,000 (Rappaport \& van den
Heuvel 1982, Meurs \& van den Heuvel 1989)\\
-- binary periods lie in the range 16d -- 400d\\
-- pulse periods lie in the range 0.07s -- 1413s\\
-- the source names are primarily based upon their coordinates.\\

\section{Evolution}

The evolutionary history of Be stars in these Be/X-ray binary systems
is somewhat different to that of their isolated colleagues. Figure 3
shows the widely accepted evolutionary path based upon conservative
mass transfer that has been developed by van den Heuvel (1983) and
Verbunt \& van den Heuvel (1995). The important consequences of the
scenario is that wide binary orbits (200 -- 600d) are produced before
the final supernova explosion. Hence, any small asymmetries in the
subsequent SN explosion will then produce the frequently observed wide
eccentric orbits.

\begin{figure}
\begin{center}
{\epsfxsize 0.6\hsize
 \leavevmode
 \epsffile{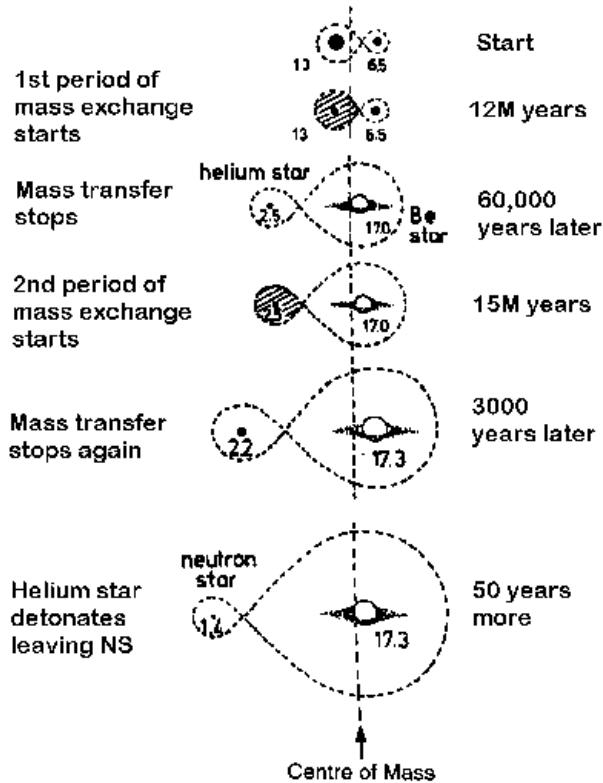}
}
\end{center}
\caption{Evolution of a Be/X-ray binary system after van den Heuvel
(1983) and others.}
\end{figure}

Of particular interest is the narrow range of spectral class of the
identified optical counterparts (see Negueruela 1998). In contrast
to the sample of isolated Be stars to be found in the Bright Star
Catalogue, there are no known Be/X-ray objects beyond spectral class
B3 - see Figure 4. Most commonly the systems have counterparts in the
B0-B2 group.

\begin{figure}
\begin{center}
{\epsfxsize 0.4\hsize
 \leavevmode
 \epsffile{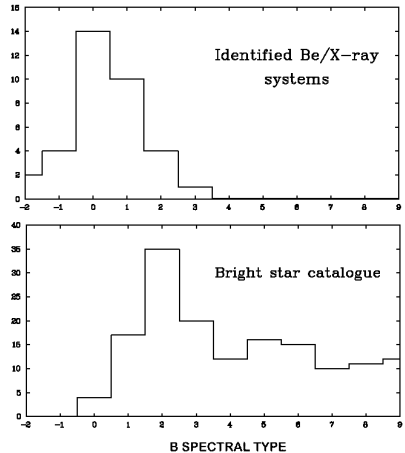}
}
\end{center}
\caption{Comparison of the spectral classes of Be stars in Be/X-ray
binaries (upper panel) with those in isolated systems (lower
panel). Diagram adapted from Negueruela (1998) with permission.}
\end{figure}

The explanation offered by Van Bever and Vanbeveren (1997) for this
phenomenom is that the wide orbits produced by the evolutionary models
are very vulnerable to disruption during the SN explosion. This will
be particularily true for the less massive objects that would make up
the later spectral classes. Hence the observed distribution is
confirming the evolutionary models.

Some evolutionary models predict the existence of other Be star +
compact object systems. In particular, the work of Raguzova and
Lipunov (1999) discusses the emergence of Be + Black Hole systems and
cites GRS 1915+105 as a possible example. Raguzova \& Lipunov (2000)
also predict that ``46\% of all Be stars formed in binary evolution
should have White Dwarf companions''. To date, however, only two B +
WD systems are known (Vennes, Berghofer \& Christian 1997 and
Burleigh \& Barstow 1999), and no Be + WD systems - but it is very
hard to detect the WD in the presence of such bright companions.

\section{Periodicities}

One of the most exciting developments in the study of Be/X-ray
binaries occurred when Corbet (1984) realised that a strong
correlation exists between the spin period of the neutron star and the
binary period of the system. Subsequently Corbet et al (1999)
developed this line of work and recently produced the more
comprehensive diagram reproduced here as Figure 5. From this figure
two things may be clearly seen: the intitial correlation for the
Be/X-ray systems is strongly confirmed, and different classes of HMXB
lie in distinctly different locations on the diagram. Hence not only
does the diagram provide a valuable tool for estimating unknown binary
periods in Be/X-ray systems, but if both periods are known then the
class of object may be deduced.

\begin{figure}
\begin{center}
{\epsfxsize 0.6\hsize
 \leavevmode
 \epsffile{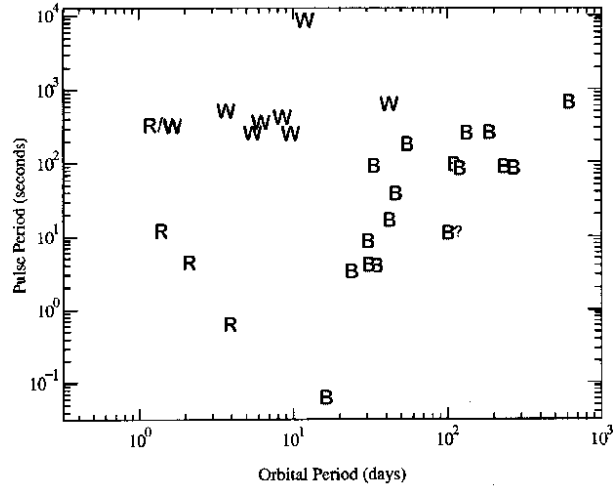}
}
\end{center}
\caption{The relationship between spin period and orbital period for
pulsating HMXBs. In the diagram B refers to Be/X-ray binaries, R to
Roche-lobe overflow systems and W to wind-fed systems. Adapted from Corbet et
al (1999) with permission.}
\end{figure}

The explanation for the striking relationship with the Be/X-ray
binaries is to be found in the process of accretion on to the neutron
star. For accretion to occur the Alfven radius must be less than the
co-rotation radius of the accreting material - otherwise the neutron
star continues to spin down until this condition is met. However, the
size of the Alfven radius depends upon the density of the surrounding
medium and larger orbits mean lower stellar wind densities in the
environment of the neutron star. Hence a relationship naturally
develops between the orbital period and the equilibrium spin period of
the neutron star.

\section{Example 1 : A1118-616}

The system A1118-616 provides an excellent example of a classic
Be/X-ray binary system. It was discovered in 1991 by chance while the
Ariel 5 satellite was observed the nearby source Cen X-3
(Eyles et al 1975). Coe et al (1994) show 
the time history of its X-ray flux over 2
months starting just before the initial outburst. During the outburst
the flux of the source increased by much more than 1 order of
magnitude and became the dominant hard X-ray source in the Centaurus
region for 7-10 days. The X-ray source then disappeared from sight for
27 years before re-emerging in the same dramatic style in 1991.

The explanation for these sudden massive X-ray outbursts lies in the
H$\alpha$ observation carried out prior to, during, and after the 1991
outburst. Note, no similar observations exist for the 1974 outburst
since the optical counterpart was not identified at that time. The
history of the H$\alpha$ observations over a 10 year period are shown
in Figure 6. From this figure it is immediately apparent why the
source went into outburst when it did - the H$\alpha$ equivalent width
had reached an exceptionally high value of more than 100\AA. 
Since the pulse period is 404s the Corbet diagram suggests
a binary period in the range 200-300d, therefore it is unlikely that
the two recorded outbursts relate to binary motion (Type I
outbursts). 
Far more likely, we are seeing two Type II outbursts, and the
normal Type I outbursts are either very minor or don't occur at
all. This could simply be due to the orbit not normally taking the neutron star
through the circumstellar disk. However, during these abnormal levels
of H$\alpha$ activity, the disk probably expands to include
the orbit of the neutron star, and hence accretion immediately
begins. Note the strikingly rapid decline of the H$\alpha$ immediately
after the outburst suggesting that the whole period of activity was
probably just due to one major mass ejection event from the Be star.

\begin{figure}
\begin{center}
{\epsfxsize 0.8\hsize
 \leavevmode
 \epsffile{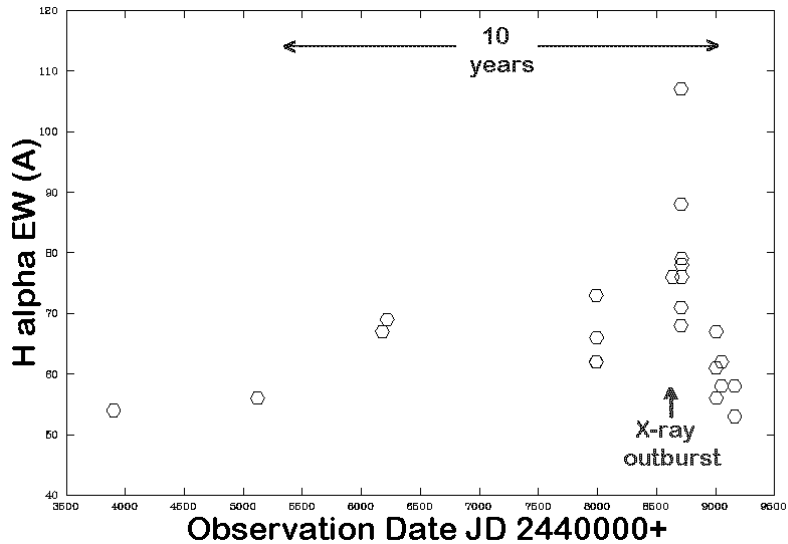}
}
\end{center}
\caption{The history of the H$\alpha$ emission from A1118-616 showing
the large rise coincident with the X-ray outburst. From Coe et al (1994)}
\end{figure}

It is interesting to compare the size of the circumstellar disk seen
in A1118-616 with that observed in other systems. Figure 7 shows a
plot of H$\alpha$ EW against the intrinsic infrared excess (also
thought to arise from free-free and free-bound emission in the
circumstellar disk). An obvious and unsurprising correlation exists
between these two parameters as is clear from the figure. The
quiescent location of A1118-616 is shown on the diagram together with
two other Be/X-ray systems. For comparison, data from a sample of
isolated Be stars (Dachs and Wamsteker 1982) is also presented. It is
interesting to note that even in quiescence, A1118-616 is at the
extreme edge of the diagram, and in fact, its peak H$\alpha$ EW value
of $\sim$110\AA ~~may be one of the very largest recorded values for
any Be star.

\begin{figure}
\begin{center}
{\epsfxsize 0.8\hsize
 \leavevmode
 \epsffile{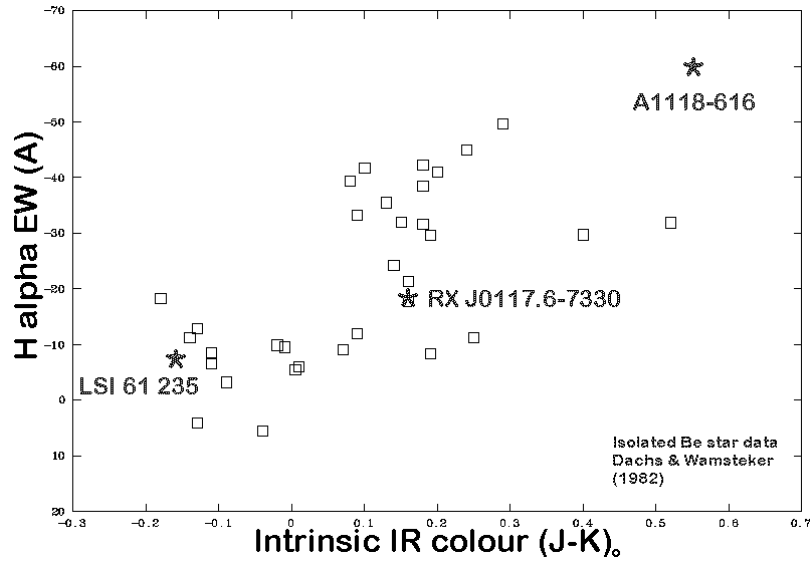}
}
\end{center}
\caption{The relationship between the intrinsic IR colours and the
H$\alpha$ equivalent width. Three Be/X-ray binary systems are compared
to a set of isolated Be stars. From Coe et al (1994)}
\end{figure}

\section{Example II : EXO2030+375}

The system EXO 2030+375 was discovered in the same manner as A1118-616
(a sudden outburst of activity from a previously unknown object), but
by EXOSAT in 1985 (Parmar et al 1989). The source was regularily
monitored by the BATSE all sky instrument on the CGRO spacecraft (see
Stollberg 1997 for an extensive discussion of this system). For long
periods of time the source has shown classic Type I outbursts every
periastron passage (binary period = 46d). Typical X-ray outbursts are
shown in Figure 8 (Norton et al, 1994).

\begin{figure}
\begin{center}
{\epsfxsize 0.4\hsize
 \leavevmode
 \epsffile{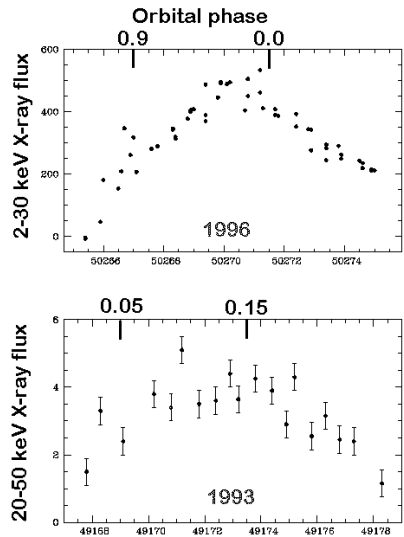}
}
\end {center}
\caption{X-ray lightcurve of two outbursts from the system EXO
2030+375. Note that the binary phase of the outburst has shifted
significantly between 1993 and 1996. From Reig \& Coe (1998) with 
permission}
\end{figure}

Then for about two years the source disappeared from the X-ray
sky and re-merged in 1996. Detailed RXTE observations (Reig \& Coe
1998) revealed a significant shift in the binary phase of the
outbursts of $\sim$0.15. This phase has since remained
unaltered, so possible precessional models of disks are ruled out. In
addition, the H$\alpha$ EW had decreased from -18\AA ~~to -8\AA,
thereby eliminating the possibility that the earlier outburst phase
arose from to an expanded circumstellar disk which the neutron star
encountered earlier in its orbit. It seems more likely that a
fundamental change has occurred in the accretion mechanism resulting
from the presence of a much smaller disk. Probably, while the disk was
large, the accretion was by Roche lobe overflow, but now it is simply
by direct stellar wind accretion. This idea is supported by the very
rapid pulse period changes seen prior to 1994 which had essentially
ceased after 1996 - the more direct coupling of Roche lobe overflow
would exert greater torques on the neutron star.

\section{Neutron star interactions with the circumstellar disk}

One of the particularily interesting questions related to studying
Be/X-ray binaries is the question of the effect of the neutron star on
the circumstellar disk. Initial work in this area by Norton et al
(1994) suggested that the dimensions of the circumstellar disk,
compared with the amount of material undergoing accretion on to the
neutron star, made it very unlikely that the presence of the neutron
star would affect the growth of the circumstellar disk.

This view has now been altered by the longer term studies carried out
by Reig, Fabregat \& Coe (1997). In Figure 9 
the maximum observed H$\alpha$ EW
is plotted against the orbital period of several systems. Though the
data are still rather sparse, there is a strong suggestion of a
correlation. The conclusion the authors draw from this diagram is that
the continually orbiting neutron star gradual erodes away the outer
edges of the circumstellar disk and inhibits its growth.

\begin{figure}
\begin{center}
{\epsfxsize 0.6\hsize
 \leavevmode
 \epsffile{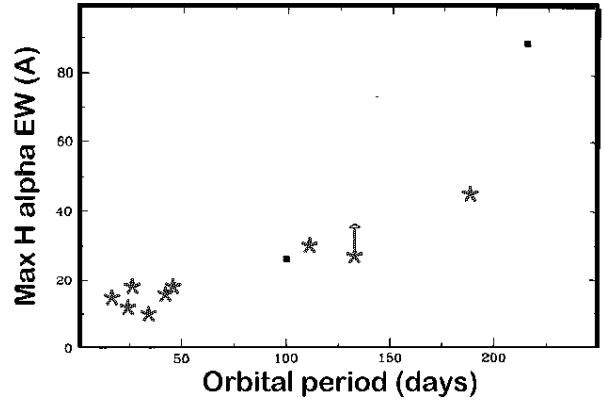}
}
\end{center}
\caption{Comparison of the orbital period with the maximum H$\alpha$
equivalent width for a variety of Be/X-ray systems. From Reig,
Fabregat \& Coe (1997) with permission.}
\end{figure}

Similarily, the same authors present a comparison of the
maximum H$\alpha$ EWs of Be stars in Be/X-ray binary systems with
isolated Be stars. Again there is a strong suggestion that, on
average, the disks are smaller in those systems with neutron star
companions. 

As yet the statistical sample sizes are small for these studies, but as more
binary periods are determined these propositions can be checked. It is
certainly important to understand how these two components in the
binary system interact.

\section{Small Magellanic Cloud}
It has come as a great surprise to discover that there are a large
number of Be/X-ray binaries in the Small Magellanic Cloud. Figure 10
shows the distribution of about half of these superimposed upon an
optical image of the SMC. 

\begin{figure}
\begin{center}
{\epsfxsize 0.6\hsize
 \leavevmode
 \epsffile{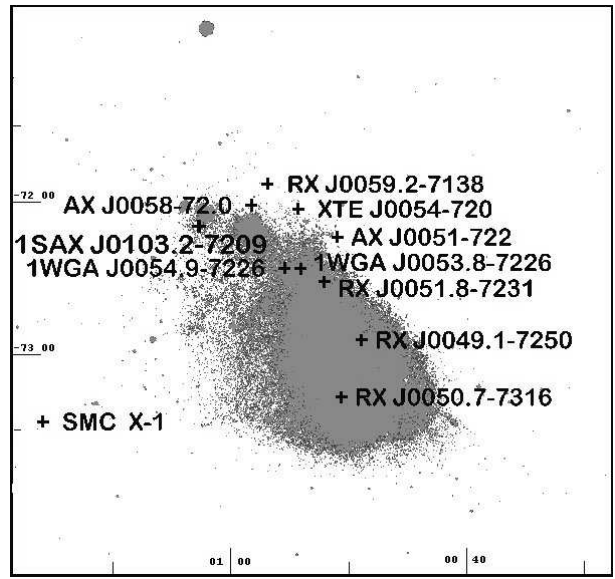}
}
\end{center}
\caption{An optical image of the Small Magellanic Cloud showing the
location of 11 out of the 20 known X-ray binary pulsar systems. From
Coe and Orosz (1999)}
\end{figure}

It is
possible to estimate the number of systems one would expect based upon
the relative masses of our galaxy and the SMC. This ratio is $\sim$50,
so with 64 known or suspected systems in our galaxy we would
only expect 1 or 2 systems in the SMC. However, Maeder, Grebel \&
Mermilliod (1999)
have shown that the fraction of Be stars to B stars is 0.39 in the SMC
compared with 0.16 in our galaxy. So this raises the expected number
of Be/X-ray systems to $\sim$3 - but we now know of 20 such
systems!

The reason for the large number of Be stars 
almost certainly lies in the history of the Magellanic
Clouds. Detailed H1 mapping by Stavely-Smith et al (1997) and Putman
et al (1998) has shown a strong bridge of material between the
Magellanic Clouds and between them and our own galaxy. Furthermore,
Stavely-Smith et al have demonstrated a the existence of a large number of
supernova remnants of a similar age ($\sim$5 Myr), strongly
suggesting enhanced starbirth has taken place as a result of tidal
interactions between these component systems. Consequently it seems
very likely that the previous closest approach of the SMC to the LMC
$\sim$100 Myrs ago may have triggered the birth of many new
massive stars which have given rise to the current population of
HMXBs. In fact, other authors (eg Popov et al 1998) claim that the
presence of large numbers of HMXBs may be the best indication of
starburst activity in a system.

Whatever the reason for the large number, the Small Magellanic Cloud
now provides us with an excellent sample of Be/X-ray systems in a
relatively compact and easily observed region of the sky.

\section{Conclusions}

Be stars in X-ray binary systems offer a whole new avenue for
exploring the Be phenomenon. The evolutionary histories are different
from the isolated systems, but their basic characteristics seem
identical. The presence of a neutron star companion provides us with
an extremely valuable probe of the circumstellar disk, and X-ray
monitoring satellites provide one of the most rapid methods for
identifying a Be star in outburst. Work in this field should continue
to make substantial contributions 
to the much broader studies of these mysterious objects.

\section{Acknowledgements}

I am very grateful to my many collaborators over the years who have
made major contributions to the topics discussed in this paper. In
particular, I wish to thank Dave Buckley (SAAO), Simon Clark (Sussex),
Juan Fabregat (Valencia), Ignacio Negueruela (Rome), Andy Norton (OU),
Pablo Reig (Crete), Paul Roche (Leicester) and Iain Steele
(Liverpool).

I also wish to thank the organisers of IAU Colloquium 175 for giving me an
opportunity to discuss the work in this field.

\section*{Discussion}

\noindent {Raguzova:} I would like to make some comments about Be stars with
white dwarfs. All white dwarfs in such systems must have very high
effective temperatures and we can detect such objects in the extreme
ultraviolet range.

\noindent {\bf Coe:} Yes, that is correct. The two known B/WD systems have been
found with EUVE - though it is proving much more difficult than one
might think. Some people believe that the numbers of such systems may
not be as large as you suggest.

\noindent {\bf Hummel:} Are the detected Be/X-ray binaries in the SMC field
stars, or members of open clusters?

\noindent {\bf Coe:} There is no evidence 
so far from the identified systems that
they are mainly found in clusters.

\end{document}